\documentclass[sigconf]{acmart}
\AtBeginDocument{%
  }

\setcopyright{acmlicensed}
\copyrightyear{2018}
\acmYear{2018}
\acmDOI{XXXXXXX.XXXXXXX}
\acmConference[Conference acronym 'XX]{Make sure to enter the correct
  conference title from your rights confirmation email}{June 03--05,
  2018}{Woodstock, NY}
\acmISBN{978-1-4503-XXXX-X/18/06}

\usepackage{enumitem}
\usepackage{comment}

\usepackage{color-edits}
\addauthor{vc}{blue}
\addauthor{sz}{purple}

\begin{document}

\title{CodingGenie: A Proactive LLM-Powered Programming Assistant}

\author{Sebastian Zhao}
\affiliation{%
  \institution{University of California, Berkeley}
  \city{Berkeley}
  \country{USA}}
\email{sebbyzhao@berkeley.edu}

\author{Alan Zhu}
\affiliation{%
  \institution{Carnegie Mellon University}
  \city{Pittsburgh}
  \country{USA}
}
\email{yixuanz2@andrew.cmu.edu}

\author{Hussein Mozannar}
\affiliation{%
  \institution{Microsoft Research}
  \city{Redmond}
  \country{USA}
}
\email{hmozannar@microsoft.com}

\author{David Sontag}
\affiliation{%
  \institution{MIT}
  \city{Cambridge}
  \country{USA}
}
\email{dsontag@mit.edu}

\author{Ameet Talwalkar}
\affiliation{%
  \institution{Carnegie Mellon University}
  \city{Pittsburgh}
  \country{USA}
}
\email{atalwalk@andrew.cmu.edu}

\author{Valerie Chen}
\affiliation{%
  \institution{Carnegie Mellon University}
  \city{Pittsburgh}
  \country{USA}
}
\email{vchen2@andrew.cmu.edu}

\renewcommand{\shortauthors}{Trovato et al.}

\begin{abstract}
While developers increasingly adopt tools powered by large language models (LLMs) in day-to-day workflows, these tools still require explicit user invocation.
To seamlessly integrate LLM capabilities to a developer's workflow, we introduce CodingGenie, a proactive assistant integrated into the code editor. 
CodingGenie autonomously provides suggestions, ranging from bug fixing to unit testing, based on the current code context and allows users to customize suggestions by providing a task description and selecting what suggestions are shown. 
We demonstrate multiple use cases to show how proactive suggestions from CodingGenie can improve developer experience, and also analyze the cost of adding proactivity. 
We believe this open-source tool will enable further research into proactive assistants. 
CodingGenie is open-sourced at \url{https://github.com/sebzhao/CodingGenie/} and video demos are available at \url{https://sebzhao.github.io/CodingGenie/}.
\end{abstract}

\ccsdesc[500]{Software and its engineering~Software notations and tools}

\keywords{coding assistants, proactive chat assistants, large language models}


\maketitle

\section{Introduction}
Developers increasingly use coding assistants powered by large language models (LLMs) in day-to-day workflows~\citep{liang2023can}, with multiple studies observing improved developer productivity and user experience~\citep{vaithilingam2022expectation,peng2023impact}.
Coding assistants (e.g., GitHub Copilot~\citep{copilot}) have typically been limited to simple code completion and refactoring~\citep{dunay2024multi} and chat conversations~\citep{xiao2023devgpt}, though recent tools like Cursor~\citep{cursor} and Windsurf Editor~\citep{Codeium} further allow for autonomous execution of tasks, including writing code, command line actions, and editing files. 
LLMs are increasingly integrated within tools for code editors for multiple purposes which are complementary to existing code completions and chat dialogues, including refactoring code~\cite{pomian2024assist}, fixing bugs~\cite{hossain2024deep}, and improving code quality~\citep{wadhwa2024core}.


Except for autocomplete suggestions, existing tools typically require explicit user invocation, i.e., the user determines when they need to use which tool and for what goals.
For example, a chat assistant requires users to decide what question to ask the assistant~\citep{ross2023programmer, chopra2023conversational} and code refactoring tools require a user to decide when to refactor functions, which is often just as important as the ability to do so~\citep{pomian2024assist}.
The influx of LLM-powered tools not only brings potential productivity benefits~\citep{eloundou2023gpts} but introduces new challenges on how to best use each tool~\citep{khurana2024and}.
While autocomplete suggestions are proactive to an extent, the next suggestion is constrained to the immediate task and context.
~\citet{chen2024need} recently proposed introducing the ability for chat assistants to provide proactive suggestions beyond code completions, finding promising results through a study on a simple web-based platform.
However, there remains a need for a proactive assistant integrated into an existing coding tool to enable broader usage and more realistic study.

\begin{figure*}[t]
  \centering
  \includegraphics[width=\linewidth]{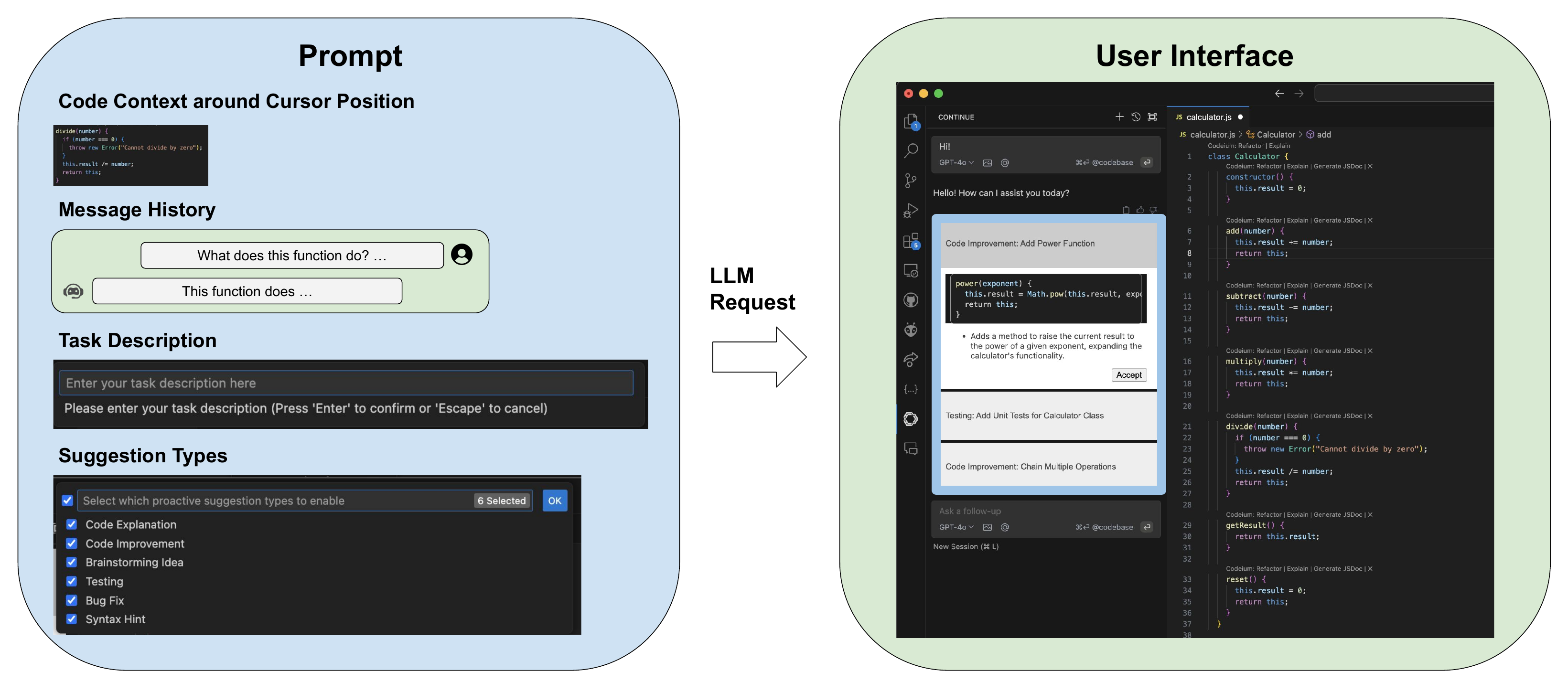}
  \caption{Overview of the CodingGenie tool. The left side of the diagram depicts the elements in a developer's workspace that make up the prompt used to generate proactive suggestions, and the right side depicts the user interface with proactive suggestions (shown in the light blue box) upon a successful generation.} 
  \Description{Overview of the CodingGenie tool. The left side of the diagram depicts the elements in a developer's workspace that make up the prompt used to generate proactive suggestions, and the right side depicts the user interface with proactive suggestions (shown in the light blue box) upon a successful generation.}
  \label{fig:overview}
\end{figure*}

In this paper, we introduce CodingGenie, an open-source implementation of proactive assistants that is easily usable and installable, built into the popular open-source coding assistant library, Continue. 
Our tool CodingGenie is named after genies that grant three wishes, as our assistant periodically provides three diverse suggestions, which include code improvements, code explanations, brainstorming ideas, additional testing, bug fixes, and syntax hints, depending on which suggestion types are enabled.
To further make CodingGenie usable in practice, we introduce two features that allow users to configure a task description and customize the types of suggestions provided.
We also discuss three use cases in which proactive suggestions may be helpful, highlighting realistic scenarios in which proactive suggestions generate helpful and directed suggestions including school assignments, resolving tickets in industry, and help with personal projects.
We include a discussion of the resource consumption of CodingGenie, and show that adopting proactive suggestions can be done at a reasonable cost.
We believe CodingGenie will allow further study of the benefits of proactivity in coding assistants, especially concerning real-world coding tasks due to its integration into VSCode.

\textit{Contributions. }  The contributions of this paper are twofold. First, we provide CodingGenie, a tool to automatically generate diverse proactive suggestions integrated into a developer's workflow with high customizability (Section~\ref{sec:tool_overview}). 
Second, we show how CodingGenie can be useful for a diverse set of use cases for a reasonable cost (Section~\ref{sec:evaluation}). 
CodingGenie is open-source at \url{https://github.com/sebzhao/CodingGenie/} and video demos are available at \url{https://sebzhao.github.io/CodingGenie/}.

\section{Tool Overview}\label{sec:tool_overview}

CodingGenie is an LLM-powered proactive assistant built into the chat interface that generates suggestions given the developer's current code context and other user input.
We describe how CodingGenie generates proactive suggestions, how developers would interact with the proactive suggestions, and how CodingGenie is implemented CodingGenie into a popular open-source coding LLM extension for VSCode, Continue.\footnote{\url{https://github.com/continuedev/continue}} 

\subsection{Generating Proactive Suggestions}

We describe how CodingGenie leverages state-of-the-art LLMs (e.g., GPT-4o~\citep{chatgpt}, Claude Sonnet-3.5~\citep{claude}), which can be selected by the user, to generate proactive suggestions.
We describe the types of suggestions that CodingGenie provides, what context the suggestions are based on, and the multiple ways that users can customize suggestions to their needs.
We overview the pieces of information that are a part of the prompt in Figure~\ref{fig:overview} (left) and provide the full prompt in the open-source repository.

\paragraph{What kinds of suggestions?} Recent work has introduced a myriad of LLM-powered tools with the ability to provide code improvements~\citep{demirci2022detecting,wadhwa2024core}, code explanations~\citep{yan2024ivie,53379}, brainstorming ideas~\citep{farah2025supporting}, unit testing~\citep{kang2023large,alshahwan2024automated}, bug fixes~\citep{jin2023inferfix,hossain2024deep}, and syntax hints~\citep{chen2024need}.
Instead of requiring users to determine when each tool should be used, we leverage state-of-the-art LLMs as a way to both proactively identify what kind of help a user might need given the current code context, which we describe next.
We view proactive suggestions as a way to unify many of the aforementioned approaches and leverage LLMs as a way to flexibly generate a diverse set of suggestions.

\paragraph{What context is included in the prompt?} The prompt contains the current code context centered around the user's cursor position,  with 500 characters of code above and below. 
The length of the code context can be adjusted based on budget and performance considerations---it is currently restricted to just the current file.
We also include the user's message history to inform which types of suggestions might be most relevant.
The prompt also includes directions to coerce the output into a format parseable, including generating a tag, a short description, code, and any explanation. 
We also include a one-shot example in the prompt, which improves adherence to the expected format.

\paragraph{How can users customize suggestions to their needs?} We introduce two ways to increase customizability, which are added to the prompt. First, users may specify a task description which is included as part of the prompt used to generate proactive suggestions. 
This enables the generation to be further aligned towards a goal. 
Furthermore, task descriptions may already be readily available and require much additional effort from the developer themselves.
For example, tickets in industry often contain a task description and further information, and can be pasted into the task description before working on the issue. 
Second, users may configure what types of suggestions are recommended based on what they are trying to accomplish, e.g. debugging versus understanding a codebase. 
Users can select from the aforementioned set of 6 categories.


\subsection{Interacting with Proactive Suggestions}

\paragraph{Interface.} Integrating proactive suggestions into a coding extension in a non-intrusive and user-friendly way also proved challenging, especially in an environment like VSCode where there are large amounts of information to process~\citep{kasatskii2023effect}. 
Due to the ever-increasing mental load on developers~\citep{rubin2016challenges}, we designed CodingGenie to interface directly into the chat panel rather than as a separate tool, as in Figure~\ref{fig:overview} (right). CodingGenie can be integrated into other popular coding assistant extensions and platforms in the same way.
To build trust in the system, ~\citet{ciniselli2023source} finds that users expect code recommendation UIs to be intuitive and easily usable. 
We found in pilot studies that proactive suggestions were confusing when they were not rendered or preserved correctly according to the user's expectations. 
As such, we choose to preserve the position of the proactive suggestions where they were recommended in the history, but add any accepted proactive suggestions to the bottom of the chat history.
These suggestions consist of three parts: the tag, which is a category that describes at a higher level what the suggestion is attempting to do; the summarized description of what the suggestion is attempting to do, which allows the user to quickly determine whether a thread is worth pursuing;  
the code and explanation, which are visible after clicking a suggestion.
Upon acceptance, the full suggestion is added to the chat interface as a message sent from the AI chatbot. 

\paragraph{Refresh Logic.} Proactive suggestions exist as a form of ``temporary'' suggestion in the chat, similar to how code completions exist as ghost text until accepted. We implement the logic for refreshing proactive suggestions as follows: Coding changes prevent any proactive suggestions for a small period of time, preventing suggestions from appearing while the user is still writing code. After a code change, a timer is started which on completion triggers a proactive suggestion request. Interacting with proactive suggestions, typing in the chatbox, and sending normal chat messages prevent any proactive suggestions for a larger period of time compared to coding changes. 
If any suggestion in a group of suggestions is accepted, the group of suggestions is kept on refresh.
Full implementation of the refresh logic is available in the open-source repository.
There is also an ability to trigger proactive suggestions manually.

\subsection{Implementation Details}

\begin{figure}[t]
  \centering
  \includegraphics[width=\linewidth]{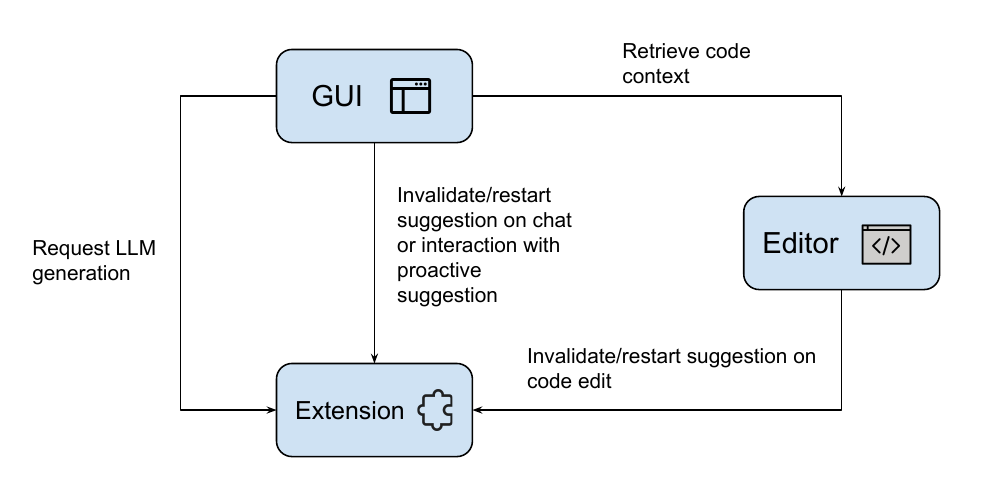}
  \caption{System design diagram for communication between the GUI, extension, and editor components of Continue to implement proactive suggestions.} 
  \Description{System design diagram for communication between the GUI, extension, and editor components of Continue to implement proactive suggestions.}
  \label{fig:system}
\end{figure}

CodingGenie is built on the open-source Continue VSCode extension, which currently supports both code autocompletion and code chat; CodingGenie exists in the code chat portion of the extension and does not interfere with any other functionality. 
We integrate proactive suggestions with the ability to specify task descriptions and suggestion types into the Continue interface.
Continue exposes a GUI and an extension that must communicate with the editor. 
For normal chat messages, this communication is only between the GUI and the extension. 
However, implementing proactive suggestions requires communication between all three components, as shown in Figure~\ref{fig:system}. 
For example, edits in the editor prevent proactive suggestions, code context is needed for proactive suggestions, the GUI must request generations from the extension, and the extension must return the result to the GUI.

\section{Evaluation}\label{sec:evaluation}

\subsection{Evaluation of Utility}
We evaluate whether CodingGenie provides helpful and relevant suggestions in three exemplary use cases.
For each use case, we request proactive suggestions at the same place in the code and report the number of relevant suggestions of each based on whether they are relevant to an overall plausible goal, specified in the use case sections below. 
We evaluate multiple CodingGenie settings and the results are summarized in Table~\ref{tab:performance}).
In both the Appendix and the video demos,\footnote{Please see \url{https://sebzhao.github.io/CodingGenie/} for video demos and further details.} we walk through each use case in detail and show examples of user interactions with CodingGenie. 

\paragraph{Personal Project.} Coding LLMs can enable novices to complete personal projects~\citep{prather2023its}.
With proactive suggestions, users may be able to easily find high-level directions to take their project directly in the editor. 
By automatically providing suggestions on what features might be interesting to add, how to refactor the code, and where bugs might occur, the user may be empowered to build projects even with limited prior experience. 
In this setting, we consider suggestions to be relevant if they present a valid feature extension of the Calculator class.

\paragraph{Industry Ticket Completion.}
In software engineering, it is common to have a ticketing system to help manage tasks. 
These tickets generally have a description of the problem, what changes may need to be done to solve the problem, and potential discussion by the ticket raiser and ticket holder on the issue.
By entering their ticket using the task description of CodingGenie, suggestions will be catered toward solving the ticket, potentially decreasing the time-until-resolution and increasing productivity. 
Relevant suggestions in this setting must attempt to parallelize inference or increase the efficiency of the benchmark.

\paragraph{Debugging School Assignment.}
Students increasingly use code autocomplete and code chat to assist them with their school assignments~\citep{kazemitabaar2023novices}. 
By configuring the types of suggestions, students may be able to receive helpful suggestions that help them learn rather than complete the assignment altogether. 
The student may only enable ``Debugging'', ``Efficiency'', and ``Improvements'' suggestions to help them improve their code quality and reason through bugs, which may assist them in learning. 
In this setting, relevant suggestions must attempt to teach or explain any completed code.

\begin{table}[t]
\centering
\begin{tabular}{@{}rccc@{}}
\toprule
\multicolumn{1}{l}{} & \multicolumn{1}{l}{Use case 1} & \multicolumn{1}{l}{Use case 2} & \multicolumn{1}{l}{Use case 3} \\ \midrule
Proactive Suggestion & 3/3 & 0/3 & 1/3 \\
+Task Description & 3/3 & 3/3 & 1/3 \\
\begin{tabular}[c]{@{}r@{}}+Type Customization\end{tabular} & 3/3 & 0/3 & 3/3 \\ \bottomrule
\end{tabular}
\caption{Number of relevant suggestions for each proactive setting across 3 different use cases: We find that customizing suggestions through task descriptions or suggestion types can increase the number of relevant suggestions.
Please see the walkthrough in the Appendix and demo videos for more information.}
\label{tab:performance}
\end{table}


\subsection{Evaluation of Cost}

While the exemplary use cases above showcase the diversity of settings where CodingGenie may boost productivity, a natural question may be the cost of using this feature, given the proactive nature of the tool.
We provide rough estimates for CodingGenie and give example numbers based on sample scenarios inferring usage information from prior work~\citep{nguyen2022empirical,jiang2024analysis}, comparing to the base costs of code completions and chat.
The base cost for autocompletion and chat can be determined based upon the frequency of requests \textit{f} and the cost per request \textit{c}. 

\begin{equation}\label{eq:cost}
\text{total cost} = f \cdot c
\end{equation}

The cost per request \textit{c} may vary based upon certain factors, including the context length and model used. Here, we use the current prices for GPT-4o for chat and Codestral for code autocomplete. 
Autocomplete models tend to be cheaper than larger chat models (e.g., as of Jan 2025, GPT-4o costs \$2.5 per million input tokens and \$10 per million output tokens and Codestral costs \$0.2 per million input tokens and \$6 per million output tokens). 
We also assume that the context length is equivalent for code autocomplete and chat, and the regular chat cost is small compared to the proactive suggestions due to shorter context lengths and less frequent use.

Assuming input tokens are much higher than output tokens, particularly for long code context, proactive requests cost roughly 10 times more than autocomplete based on Equation~\ref{eq:cost}. 
Given that proactive suggestions are only suggested after a delay in typing, proactive suggestions appear strictly less than autocomplete suggestions. 
In the worst case where a coder waits long enough for both autocomplete and proactive suggestions to appear, the final cost is roughly 11x without proactivity. 
On the other hand, if the user only sees proactive suggestions around 1/10 of the time as autocomplete suggestions, the final cost is only around 2x the cost---given the larger delay until requesting proactive suggestions and more strict refresh logic, we predict proactive suggestions should be requested at least an order of magnitude less than autocomplete, but further study is necessary to verify this.
As such, as of January 2025, GitHub Copilot costs \$10 per month for individual users, so we estimate adoption of proactive suggestions to be upper bounded by \$20 per month.
Tuning the contribution of proactive suggestions to cost is further possible by tuning the length of code context available and refresh logic.  

\section{Related Work}


\textbf{LLM-powered coding tools.} LLMs are increasingly integrated within various tools in the software development life cycle; these tools can provide code improvements~\citep{demirci2022detecting,wadhwa2024core}, code explanations~\citep{yan2024ivie,53379}, brainstorming ideas~\citep{farah2025supporting}, unit testing~\citep{kang2023large,alshahwan2024automated}, bug fixes and solving project issues ~\citep{jin2023inferfix,hossain2024deep,10.1145/3650212.3680384}, and syntax hints~\citep{nam2024using}.
For example,~\citet{wadhwa2024core} utilizes a proposer and ranker LLM to generate code quality revisions. 
In another tool,~\citet{lemieux2023codamosa} helps increase test coverage by utilizing LLMs to direct testing towards under-covered functions. 
Instead of focusing on a specific capability, we instead leverage LLMs as a way to proactively identify what kind of help a user might need, instead of requiring them to determine when each tool should be invoked.
We incorporate user context into the prompt to the LLM and additional user customization to encourage more relevant suggestions.

\textbf{Proactivity.} The most well-known and widely used proactive tool for software development is in-line code completions (e.g., GitHub Copilot~\cite{copilot})~\citep{liang2023large}.
Aside from proactive code completions, a few related works have studied proactive assistance for targeted goals of fixing error messages, understanding LLM generations and outcomes of AI-generated code~\cite{errormessage, feedback, ahmed2020characterizing,liveprogramming1}.
CodeAid~\cite{10.1145/3613904.3642773} allows for tailored suggestions in a structured interface with a chat-based agent focused on educational purposes, and also considers design decisions like the degree of proactivity. 
In contrast, focusing on more general-purpose coding agents, ~\citet{chen2024need} conducted a preliminary study in a web-based coding environment, demonstrating the potential productivity benefits of a more general-purpose proactive assistant.
In this work, we extend some of these initial ideas with further user customizations and contribute an open-source implementation of proactive assistants for real-world coding tasks---CodingGenie---to enable practical studies of proactive assistants.

\section{Conclusion}
We proposed and released a proactive assistant tool, CodingGenie, integrated into VSCode. 
This proactive assistant has the ability to autonomously provide chat suggestions based upon context, enabling high level suggestions with knowledge of code context, as well as a high degree of customizability configuring the task description and types of suggestions shown.
We discuss several use cases and show how CodingGenie can provide relevant suggestions for a diverse set of use cases 
We also analyze the cost of adding proactivity, and find it increases the cost of using CodingGenie, but may be tuned to reduce the impact. 
We believe CodingGenie can enable further research into proactive assistants.

\bibliographystyle{ACM-Reference-Format}
\bibliography{main}

\end{document}